\documentstyle[12pt]{article}

\newcommand{\MS}{\overline{MS}}

\begin{document}

\setcounter{page}{0} \thispagestyle{empty}

\begin{flushright} {\bf 
DESY 00-059 \\
hep-ph/0004008
} \end{flushright}

\vskip 2cm

\begin{center}
{\Large {\bf NLO corrections to the BFKL equation in QCD and in 
supersymmetric gauge theories. 
}} \\[0pt]
\vspace{1.5cm} {\large \ A.V. Kotikov\footnote{
Alexander von Humboldt fellow. On leave of absence from the Particle 
Physics Laboratory of the Joint Institute for Nuclear Researches, 
141980 Dubna, Russia}
}
\\[0pt]
\vspace{0.5cm} {\em II Institut f\"{u}r Theoretische Physik\\[0pt]
Universit\"{a}t Hamburg\\[0pt]
Luruper Chussee 149 \\[0pt]
22761 Hamburg, Germany
}\\[0pt]

\vspace{0.5cm} and \\[0pt]
\vspace{0.5cm} {\large \ L.N. Lipatov\footnote{
Supported by the
INTAS grant 97-31696 and by the Deutsche Forschungsgemeinschaft}
} 
\\[0pt]
\vspace{0.5cm} {\em St. Petersburg State University and \\[0pt]
Petersburg Nuclear Physics Institute  \\[0pt]
Gatchina, Orlova Roscha, 188300, St. Petersburg, Russia
}\\[0pt]
\end{center}

\vspace{1.5cm}\noindent

\begin{center}
{\bf Abstract}
\end{center}

We study next-to-leading corrections to the integral kernel of the BFKL
equation for high energy cross-sections in QCD and in supersymmetric gauge
theories. The eigenvalue of the BFKL kernel is calculated in an analytic form
as a function of the anomalous
dimension $\gamma $ of the local gauge-invariant operators and their
conformal spin $n$. For the case of an
extended $N=4$ SUSY the kernel is significantly simplified. In particular,
the terms non-analytic in $n$ are canceled. We discuss the relation between 
the DGLAP and BFKL equations in the $N=4$ model.\\


\newpage


\pagestyle{plain}

\section{Introduction}

\indent

The Balitsky-Fadin-Kuraev-Lipatov (BFKL) equation \cite{BFKL, BL} is used
now together with the Dokshitzer-Gribov-Lipatov-Altarelli-Parisi (DGLAP)
equation \cite{DGLAP} for a theoretical description of structure functions
of the deep-inelastic $ep$ scattering at small values of the Bjorken
variable $x$. In this kinematical region the structure functions are
measured by the H1 and ZEUS collaborations \cite{H1} at DESY. For the DGLAP
equation the radiative corrections to the splitting kernels are well known
\cite{corAP}. Although the BFKL equation in the leading logarithmic
approximation (LLA) was constructed many years ago, the calculation of the
next-to-leading corrections to its kernel was started only in 1989 
\cite{LF89}
and completed recently \cite{FL, CaCi}. The total cross-section for the
colourless particle scattering is expressed in terms of the solution $%
G_{\omega }(\overrightarrow{q},\overrightarrow{q^{\prime }})$ of the BFKL
equation for the $t$-channel partial wave $f_j (\overrightarrow{q} 
\overrightarrow{q^{\prime }};t)$ ($j=1+\omega ,\,\, t=0$) describing the 
scattering of the
reggeized gluons having their initial and final transverse momenta $%
\overrightarrow{q}$ and $\overrightarrow{q^{\prime }}$ correspondingly.

For unpolarized colourless particle collisions one can average the 
BFKL kernel over the
angle between the two-dimensional vectors $\overrightarrow{q},%
\overrightarrow{q^{\prime }}$. The eigenvalue of the averaged 
kernel was calculated in \cite{FL}. The high energy behaviour of the 
total cross-sections $\sigma  
\sim s^{\Delta }$ is related with the pomeron intercept $\Delta $. 
It turns out, that in $\overline{MS}$%
-renormalization scheme the NLO corrections to  $\Delta $ 
are negative and big. However, with the use of non-abelian
renormalization schemes and the BLM procedure for the optimal
scale setting it was shown \cite{bfklp}, that the BFKL predictions for the
total cross-section of the process $\gamma ^{*}\gamma ^{*}\rightarrow
hadrons $ turn out to be in an agreement with the experimental data obtained
by the L3 collaboration
 \cite{L3}. There are other resummation methods (see \cite
{resum} and references therein) which lead to quite similar numerical
results.

Because the initial virtual photons of the process $\gamma ^{*}\gamma
^{*}\rightarrow hadrons$ can be polarized, one should know also 
the NLO corrections to the spin correlators at high energies. It
is one of the reasons, why in this paper we calculate  
next-to-leading
corrections to the intercept $\Delta $ of the BFKL pomeron with an arbitrary
conformal spin $n$. The case $n=0$ corresponds to the cross-section $\sigma
_{0}$ averaged over the angle $\vartheta $ between the photon spins (see \cite
{FL}) and the case $n=2$ describes the high energy behaviour $s^{\Delta
_{2}} $ of the contribution $\sigma _{2}$ proportional to $\cos
^{2}\vartheta -\frac{1}{2}$.

On the other hand, the contribution of higher conformal spins is interesting
from a pure theoretical point of view, because it is related to remarkable
mathematical properties of the BFKL equation. To begin with, this equation
in LLA is invariant under the M\"{o}bius transformation of holomorphic
impact parameters

\begin{equation}
\rho _{k}\rightarrow \frac{a\rho _{k}+b}{c\rho _{k}+d}\,,
\end{equation}
for arbitrary complex values of $a,\,b,\,c$ and $d$ \cite{conf}. This
invariance allows us to find its solution for arbitrary momentum transfers
$q=\sqrt{-t}$ providing that we know the eigenvalues $\Delta (m,%
\widetilde{m})$ of the kernel for all conformal spins $n$. The conformal
weights $m$ and $\widetilde{m}$ are related to the anomalous dimension $%
\gamma =\frac{1}{2}+i\nu $ of the local operators belonging to a basic
series of unitary representations of the M\"{o}bius group as follows

\begin{equation}
m=\gamma +\frac{n}{2}\,,\,\,\widetilde{m}=\gamma -\frac{n}{2}\,.
\end{equation}
Further, in LLA the homogeneous BFKL equation in the impact parameter
representation can be written as the Schr\"{o}dinger equation

\begin{equation}
H_{12}\Psi =E_{12}\Psi \,,\,\,\Delta =-\frac{g^{2}}{8\pi ^{2}}
\min{E_{12}}\,,
\end{equation}
where $\Delta $ is the BFKL intercept. The pair Hamiltonian $H_{kl}$ for the
interaction of two reggeized gluons with their holomorphic coordinates $\rho
_{k}$, $\,\rho_{l}$ and momenta $p_{k}=i\partial _{k}$, $p_{l}=i\partial
_{l}$ can be presented as a sum of the holomorphic and anti-holomorphic
Hamiltonians \cite{integr}:

\begin{equation}
H_{kl}=h_{kl}+h_{kl}^{*}\,.
\end{equation}
Here

\begin{equation}
h_{kl}=\ln (p_{k}p_{l})+\frac{1}{p_{k}}\ln (\rho
_{kl})p_{k}+\frac{1}{p_{l}}%
\ln (\rho _{kl})p_{l}+2\gamma \,\,,
\end{equation}
$\gamma $ is the Euler constant and $\rho _{kl}=\rho _{k}-\rho _{l}$.

The holomorphic separability of the Hamiltonian is very important for 
the
solution of the generalized BFKL equation \cite{BKP} for a composite state
of $n$ reggeized gluons:
\begin{equation}
H\Psi =E\Psi \,,\,\,\Delta _{n}=-\frac{g^{2}}{8\pi ^{2}}E\,.
\end{equation}
The Hamiltonian $H$ in the multi-colour QCD $N_{c}\rightarrow \infty $ can
be presented as follows \cite{integr}:
\begin{equation}
H=\frac{1}{2}(h+h^{*}),\,\,h=\sum_{k=1}^{n}\,h_{k,k+1}\,,\,\,\,
[h,h^*]=0 \,.
\end{equation}
Due to the commutativity of $h$ and $h^{*}$ the function $\Psi $ has the 
property of the holomorphic factorization \cite{integr} 
\begin{equation}
\Psi =\sum_{r<s}\,C_{rs}\,\psi _{r}\,\widetilde{\psi } _{s}\,,
\end{equation}
where $\psi _{r}$ and $\widetilde{\psi _{s}}$ are solutions of the
Schr\"{o}dinger equations 
\begin{equation}
h\psi _r=\varepsilon \psi _r \,,\,\,h^{*}\widetilde{\psi 
}_s=\widetilde{\varepsilon
}\widetilde{\psi }_s\,,\,\,E=\frac{1}{2}(\varepsilon
+\widetilde{\varepsilon }%
)\,
\end{equation}
with the same values of energies
$\varepsilon $ and $\widetilde{\varepsilon }$  in the
holomorphic and anti-holomorphic subspaces correspondingly.
The coefficients $C_{rs}$ should be chosen from the requirement of a
single-valuedness of $\Psi $ in the two-dimensional impact parameter space.

The holomorphic Hamiltonian $h$ has the integrals of motion $q_{r}$ ($%
r=1,2...n$) \cite{integr}
\begin{equation}
q_{r}=\sum_{i_{1}<i_{2}<...<i_{r}}\rho _{i_{1}i_{2}}\rho
_{i_{2}i_{3}}...\rho _{i_{r}i_{1}}\,p_{i_{1}}p_{i_{2}}...p_{i_{r}}
\end{equation}
with the properties
\begin{equation}
\lbrack q_{r},q_{s}]=0\,,\,\,[q_{r},h]=0\,.
\end{equation}
Moreover, it coincides with the local Hamiltonian of an integrable 
Heisenberg spin model \cite{LFK}.

The integrability of the model is related to a duality symmetry
\cite{dual}
of $h$ and  $q_{r}$ under the
canonical transformation
\begin{equation}
p_{r}\rightarrow \rho _{r,r+1}\rightarrow p_{r+1}
\end{equation}
combined with the transposition of these operators.

In particular, in the case of the compound state of
three reggeized gluons, where we can use the conformal anzatz
\begin{equation}
\Psi _{m,\widetilde{m}}=\left( \frac{\rho _{23}}{\rho _{20}\rho _{30}}%
\right) ^{m}\,\left( \frac{\rho _{23}^{*}}{\rho _{20}^{*}\rho _{30}^{*}}%
\right) ^{\widetilde{m}}\varphi _{m,\widetilde{m}}(x,x^{*})\,,\,\,x=\frac{%
\rho _{12}\rho _{30}}{\rho _{10}\rho _{32}}\,,
\end{equation}
the duality equation for $\varphi _{m,\widetilde{m}}(x,x^{*})$ can be
written in the pseudo-differential form
\begin{equation}
Q_{m,\widetilde{m}}\varphi _{m,\widetilde{m}}=a_{m}\,a_{\widetilde{m}%
}\,\varphi _{m,\widetilde{m}}=\left| \lambda _{m}\right| \,\varphi _{1-m,1-%
\widetilde{m}}\,,\,\,a_{m}=x(1-x)\,(i\partial )^{1+m}.\,
\end{equation}
Using this equation, 
new three-gluon states having the pomeron and  odderon
quantum numbers and the intercept $\Delta =0$ were
constructed \cite{bvl}. Note, that the effective action describing the
interaction of reggeized gluons with quarks and gluons  was
derived in ref. \cite{Lipatov}.

To investigate a possibility of the holomorphic separability of the BFKL
kernel at the next-to-leading approximation in QCD and in extended
supersymmetric gauge theories is another reason for the calculation of the
$n
$-dependence of the pomeron intercept.  A presumable integrability of the
BFKL dynamics in the $N=4$ supersymmetric field theory at $N_{c}\rightarrow
\infty $ could be related to remarkable properties of the renormalization
group equations in this model. It is well known, that the $\beta $-function
here is zero in all orders of the perturbation theory. Moreover, if we
consider conformally invariant evolution equations for  matrix elements
of the quasi-partonic operators introduced in ref. \cite{qp}, it turns out
\cite{N=0}, that for $N=4$ the pair anomalous dimensions for the
supersymmetric twist-$2$ partonic operators with the Lorentz spin $j$ are
proportional to $\psi (j-1)-\psi (1)$. It means, that for $N_{c}\rightarrow
\infty $ the corresponding Hamiltonian coincides with the local
Hamiltonian for
an integrable Heisenberg spin model similar to that obtained for the
reggeized gluons (see \cite{integr, LFK}). The integrability of the
evolution equations at the $N=4$ theory in the multi-colour limit is
important for a verification of the Maldacena guess \cite{Malda}.

In this paper we extend the analysis of article \cite{BL} to the NLO level.
Namely, we calculate eigenvalues of the
angle-dependent eigenfunctions of the
NLO BFKL equation for the $t$-channel partial waves in the framework of QCD
and in gauge models with an extended supersymmetry. In the case of QCD this 
analysis
allows us to find main NLO corrections to an angle-depended part of
cross-sections. On the basis of the study of gauge models with an extended
supersymmetry one can attempt to find a possible source for the violation of
the M\"{o}bius symmetry, which is important for the calculation of the $t$%
-dependent BFKL-kernel in QCD at the NLO level.\newline

The article is organized as follows. In Section 2 we review the
results of ref. \cite{FL},  
calculate the eigenvalues of the NLO BFKL equation in QCD and discuss 
the asymptotic behaviour of cross-sections at large $s$. Section 3
contains the analysis of the gluino and scalar particle contributions and
our results for eigenvalues of the NLO BFKL equation in supersymmetric
theories. A summary is given in Conclusion. In Appendix A we discuss the
reasons for the appearance of the special function ${\rm Ls}_{3}(x)$ in our
formulae
and in Appendix B we consider the Pomeranchuck singularity in QED 
and in its supersymmetric extension.


\section{NLO corrections to BFKL kernel in QCD}

\indent

To begin with, we review shortly the results of ref.\cite{FL}, where the QCD
radiative corrections to the BFKL integral kernel at $t=0$ were calculated.
We discuss only the formulae important for our analysis.

The total cross-section $\sigma (s)$ for the high energy scattering of
colourless particles $A,B$ written in terms of their impact factors $\Phi
_{i}(q_{i})$ and the $t$-channel partial wave $G_{\omega }(q,q^{\prime })$
for the gluon-gluon scattering is

\begin{equation}
\sigma (s)~=~\int \frac{d^{2}q\,d^{2}q^{\prime }}{(2\pi
)^{2}\,q^{2}\,q^{\prime 2}}\Phi _{A}(q)\,\Phi _{B}(q^{\prime
})\int_{a-i\infty }^{a+i\infty }\frac{d\omega }{2\pi i}\left( 
{\frac{s}{s_0}}\right) ^{\omega }G_{\omega }(q,q^{\prime }), ~~~
s_0 = |q||q^{\prime }|.
\end{equation}

Here $q$ and $q^{\prime }$ are transverse momenta\footnote{%
To simplify equations hereafter (except Appendix B) we omit arrows
in the notation of transverse momenta $\overrightarrow{q},~ \overrightarrow{%
q^{\prime}},~ \overrightarrow{q_1},~\overrightarrow{q_2},~ ...~$, i.e. in
our formulae the momenta $\overrightarrow{q},~
\overrightarrow{q^{\prime}},~%
\overrightarrow{ q_1},~\overrightarrow{q_2},~ ...~$ will be represented as
$q,~q^{\prime},~q_1,~q_2,~...~$, respectively. Note, however, that the
momenta $p_A$ and $p_B$ are $D$-space momenta.} of virtual gluons and 
$s=2p_{A}p_{B}$ is the squared invariant mass for the colliding particle
momenta $p_{A}$ and $p_{B}$.

Using the dimensional regularization in the $\overline{MS}$-scheme to
 remove ultraviolet and infrared divergences in
intermediate expressions, the BFKL equation for $G_{\omega }(q,q^{\prime })$
can be written in the following form
\begin{equation}
\omega G_{\omega }(q,q_{1})~=~\delta ^{D-2}(q-q_{1})+\int
d^{D-2}q_{2}\,K(q,q_{2})\,G_{\omega }(q_{2},q_{1})\,,
\end{equation}
where

\begin{equation}
K(q_{1},q_{2})~=~2\,\omega (q_{1})\,\delta
^{D-2}(q_{1}-q_{2})+K_{r}(q_{1},q_{2})
\end{equation}
and the space-time dimension $D=4+2\varepsilon $ for $\varepsilon \to 0$.
The gluon Regge trajectory $\omega (q)$ and the integral kernel $%
K_{r}(q_{1},q_{2})$ related to the real particle production have been
calculated in \cite{LF89}, \cite{FFK}, \cite{FL93}, \cite{FL96} and \cite
{CCF}. The final results (see \cite{FL}) can be written in the following
form correspondingly for the Regge trajectory
\[
\omega (q)=-\overline{\alpha }_{\mu }\left( \frac{2}{\varepsilon }+2\ln
\left( \frac{q^{2}}{\mu ^{2}}\right)
\right) -~\overline{\alpha }_{\mu }^{2}%
\Biggl[ \left( \frac{11}{3}-\frac{2}{3}\frac{n_{f}}{N_{c}}\right)
\left( 
\frac{1}{\varepsilon ^{2}}-\ln ^{2}\left( \frac{q^{2}}{\mu ^{2}}\right)
\right) 
\]
\begin{equation}
+\left( \frac{67}{9}-2\zeta (2)-\frac{10}{9}\frac{n_{f}}{N_{c}}\right)
\left( \frac{1}{\varepsilon }+2\ln \left( \frac{q^{2}}{\mu ^{2}}\right)
\right) -\frac{404}{27}+2\zeta (3)+\frac{56}{27}\frac{n_{f}}{N_{c}}\Biggr]
\end{equation}
and for the kernel 
\[
K_{r}(q_{1},q_{2})=\frac{4\overline{\alpha }_{\mu }\mu ^{-2\varepsilon }} {%
\pi ^{1+\varepsilon }\Gamma (1-\varepsilon )}\Biggl\{ \frac{1} {%
(q_{1}-q_{2})^{2}}\Biggl(1 
\]
\[
+\overline{\alpha }_{\mu }\Biggl[\left( \frac{11}{3}-\frac{2}{3} \frac{n_{f}%
}{N_{c}}\right) \frac{1}{\varepsilon }\left( 1-\left( {\frac{%
(q_{1}-q_{2})^{2}}{\mu ^{2}}}\right) ^{\varepsilon } \left( 1-\zeta
(2)\varepsilon ^{2}\right) \right) -{\ln }^{2}\left( {\frac{q_{1}^{2}}{%
q_{2}^{2}}}\right) 
\]
\[
+\left( \frac{(q_{1}-q_{2})^{2}}{\mu ^{2}}\right) ^{\varepsilon } \left( 
\frac{67}{9}-2\zeta (2)-\frac{10}{9}\frac{n_{f}}{N_{c}}+\varepsilon \left( -%
\frac{404}{27}+14\zeta (3)+\frac{56}{27}\frac{n_{f}}{N_{c}} \right) \right) %
\Biggl] \Biggr)
\]
\[
+\overline{\alpha }_{\mu }\Biggl[\left( 1+\frac{n_{f}}{N_{c}^3}\right) \frac{%
3(q_{1}q_{2})^{2}-2q_{1}^{2}q_{2}^{2}}{16q_{1}^{2}q_{2}^{2}} \left( \frac{2}{%
q_{1}^{2}}+\frac{2}{q_{2}^{2}}+ \left( \frac{1}{q_{2}^{2}}-\frac{1}{q_{1}^{2}%
} \right) {\ln }\left( {\frac{q_{1}^{2}}{q_{2}^{2}}}\right) \right)
\]
\[
-\left( 3+\left( 1+\frac{n_{f}}{N_{c}^3}\right) \left( 1-\frac{
(q_{1}^{2}+q_{2}^{2})^{2}}{8q_{1}^{2}q_{2}^{2}}-\frac{ 2q_{1}^{2
}q_{2}^{2}-3q_{1}^{4}-3q_{2}^{4}}{16q_{1}^{4}q_{2}^{4}} (q_{1}q_{2})^{2}%
\right) \right) \int_{0}^{\infty }\frac{dx}{ q_{1}^{2}+xq_{2}^{2}}\ln \left|
\frac{1+x}{1-x}\right|
\]
\[
+ \frac{(q_{1}^{2}-q_{2}^{2})}{(q_{1}-q_{2})^{2}(q_{1}+q_{2})^{2}}\left( \ln
\left( \frac{q_{1}^{2}}{q_{2}^{2}}\right) \ln \left( 
\frac{q_{1}^{2}q_{2}^{2}(q_{1}-q_{2})^{4}}
{(q_{1}^{2}+q_{2}^{2})^{4}}\right) -2
{\rm Li}_{2}\left( -\frac{q_{1}^{2}}{q_{2}^{2}}\right) 
+2{\rm Li}_{2}\left( 
\frac{q_{1}^{2}}{q_{2}^{2}}\right) \right) 
\]
\begin{equation}
-\left( 1-\frac{(q_{1}^{2}-q_{2}^{2})}{(q_{1}-q_{2})^{2}(q_{1}+q_{2})^{2}}
\right) \left( \int_{0}^{1}-\int_{1}^{\infty }\right) \frac{dx}{
(q_{2}-xq_{1})^{2}}\ln \left( \frac{xq_{1}^{2}}{q_{2}^{2}}\right) \Biggr] %
\Biggr\},
\end{equation}
where ${\rm Li}_{2}(x)$ and $\zeta (n)$ are the Euler dilogarithm and
Riemann $\zeta $-functions: 
\begin{equation}
{\rm Li}_{2}(x)=-\int_{0}^{x}\frac{dt}{t}\ln (1-t),~~~
\zeta(n)=\sum_{k=1}^{\infty }k^{-n}.
\end{equation}
Here we introduced also 
\begin{equation}
\overline{\alpha }_{\mu }=\frac{g_{\mu }^{2}N_{c}\Gamma (1-\varepsilon )}{
(4\pi )^{2+\varepsilon }}
\end{equation}
for the colour group $SU(N_{c})$ and $g_{\mu }$ is the QCD coupling constant
fixed at the normalization point $\mu $ in the $\overline{MS}$-scheme.

As it was shown in \cite{BL}, a complete and orthogonal set of
eigenfunctions of the homogeneous BFKL equation in LLA is 
\begin{equation}
G_{n,\gamma }(q/q^{\prime },\theta )~=~\left( \frac{q^{2}}{q^{\prime 2}}
\right) ^{\gamma -1}e^{in\theta }
\end{equation}

The BFKL kernel in this representation is diagonalized up to the effects
related with the running coupling constant $\alpha _{s}(q^{2})$: 
\begin{equation}
\omega =\frac{\alpha _{s}(q^{2})N_{c}}{\pi }\biggl[ \chi (n,\gamma )+\delta
(n,\gamma )\frac{\alpha _{s}(q^{2})N
_{c}}{4\pi }\biggr] \,.
\end{equation}
The calculation 
of the correction $\delta (n,\gamma )$ for all conformal spins
is one of main results of our paper.



\subsection{The technique of calculations}

\indent

The integration over the angle $\theta $ of the transverse
gluon momentum has the form

\[
\int_{-\pi }^{\pi }d\theta e^{in\theta }=2\int_{0}^{\pi }d\theta \cos
(n\theta )=2\int_{0}^{\pi }d\theta {\rm T}_{n}(\cos \theta ),
\]
where ${\rm T}_{n}(\cos \theta )$ are the Chebyshev polynomials
(below $n$ is a positive integer number).

In the framework of the dimensional regularization a natural extension of
the Chebyshev polynomial is the Gegenbauer polynomial ${\rm C}_{n}^{\lambda
}(\cos \theta )$ (see \cite{DuFiSi}-\cite{Kotikov96}) with the index $%
\lambda $ related to the transverse space dimension 
$D-2$ as $\lambda =(D-2)/2-1=D/2-2$. When $\lambda =0$ we return to the
Chebyshev polynomials because

\begin{eqnarray}
{\rm C}_n^0 (\cos \theta) \equiv \lim_{\lambda \to 0} \frac{1}{\lambda} {\rm
C}_n^{\lambda} (\cos \theta) = \frac{2}{n} {\rm T}_n (\cos \theta).
\nonumber
\end{eqnarray}

To calculate the terms singular at $\varepsilon \to 0$ it is convenient to
to use the formulae from \cite{CheKaTka, Kotikov96} for the traceless
products 

\begin{eqnarray}
x^{(\mu_1,...,\mu_n)}~=~
\sum_{p \geq 0}^{[n/2]} \frac{n!(-1)^{p} \Gamma(n-p+\lambda)}
{2^{2p} p! (n-2p)!\Gamma(n+\lambda)}~
g^{\mu_1\mu_2}...g^{\mu_{2p-1}\mu_{2p}}~x^{2p}~x^{\mu_{2p+1}}...x^{\mu_n}\,.
 \nonumber 
\end{eqnarray}
The traceless products are related to the Gegenbauer polynomials 
${\rm C}_{n}^{\lambda}(\cos \theta )$ as follows

\[
\frac{n}{2}\frac{1}{\lambda }{\rm C}_{n}^{\lambda }(\hat{q}_{1}\hat{q}%
_{2})=S_{n}^{(1)}(\lambda )\frac{q_{1}^{(\mu _{1},...,\mu _{n})}q_{2}^{(\mu
_{1},...,\mu _{n})}}{{(q_{1}^{2}q_{2}^{2})}^{n/2}},
\]
where the symbols $\hat{q}_{1}$ and $\hat{q}_{2}$ are the 
unit vectors parallel to
the transverse momenta $q_{1}$ and $q_{2}$,
\[
\hat{q}_{1}\hat{q}_{2}=\frac{(q_{1}q_{2})}{{(q_{1}^{2}q_{2}^{2})}^{1/2}}
\]
and $S_{n}^{(1)}(\lambda )$ is the following factor
\[
S_{n}^{(1)}(\lambda )=\frac{2^{n-1}\Gamma (n+\lambda )}{\Gamma (n)\Gamma
(1+\lambda )}
\]

After the integration over the momentum $q_{2}$ in Eq.(16) 
we obtain the product of the same vectors \cite{CheKaTka, Kotikov96}
\[
\frac{q_{1}^{(\mu _{1},...,\mu _{n})}q_{1}^{(\mu _{1},...,\mu _{n})}}{%
q_{1}^{2n}}=\frac{\Gamma (\lambda )\Gamma (n+2\lambda )}{2^{n}\Gamma
(2\lambda )\Gamma (n+\lambda )}=S_{n}^{(2)}(\lambda )
\]

Then in the front of the expressions, obtained from the
contributions singular at $\varepsilon \to 0$, the following coefficient
\[
S_{n}(\lambda )=S_{n}^{(1)}(\lambda )S_{n}^{(2)}(\lambda )=\frac{\Gamma
(n+2\lambda )}{\Gamma (n)\Gamma (1+2\lambda )}
\]
appears. Because $S_{n}(\lambda )$ is factorized 
and $S_{n}(0)=1$, it does not have any influence on our results.
Thus, the contributions singular at $\varepsilon \to 0$ can be calculated 
easily with the use of
the formulae from \cite{CheKaTka, Kotikov96}.\newline

The $\varepsilon $-independent terms are obtained in a similar way, because
one can use of the well known formulae (see \cite{Kotikov96}) to 
calculate the
contributions of the Feynman diagrams containing $\theta $-functions
\[
\theta(y)=
\left\{ 
\begin{array}{rl}
1, & \mbox{ if }y\geq 0 \\ 
0, & \mbox{ if }y<0
\end{array}
\right. 
\]
in the integrands. These $\theta $-functions come from an expansion of the
expressions\\ $(q_1+q_2)^{-2},~(q_1-xq_2)^{-2}$ and $\ln(q_1-q_2)^2$  
in the r.h.s. of eq.(19) (see eq.(\ref{6.1}),
for example).

However, it is more convenient to work directly with the expansion of
propagators over the Chebyshev polynomials and to take 
initially 
integrals over the angle
in the r.h.s. of eq.(19). The useful formulae for the expansion of
propagators and the properties of the Chebyshev polynomials are given below

\begin{eqnarray}
&&\hspace*{-1cm} \frac{q_1^2-q_2^2}{(q_1-q_2)^2} ~=~ \Biggl[ 1 +
2\sum^{\infty}_{n=1} {\rm T} _n (\hat q_1
 \hat q_2) {\biggl(\frac{q_1^2}{%
q_2^2} \biggr)}^{n/2} \Biggr] \theta (q_2^2-q_1^2) - \Biggl[ q_1
\leftrightarrow q_2 \Biggr] \theta (q_1^2-q_2^2)\,,  \label{6.1} \\
&&\hspace*{-1cm} \ln(q_1-q_2)^2 ~=~ \Biggl[\ln q_2^2 - 2\sum^{\infty}_{n=1}
\frac{1}{n}
{\rm T}_n (\hat q_1 \hat q_2) {\biggl(\frac{q_1^2}{q_2^2} \biggr)}^{n/2} %
\Biggr] \theta (q_2^2-q_1^2) + \Biggl[ q_1 \leftrightarrow q_2 \Biggr] %
\theta (q_1^2-q_2^2)\,,  \nonumber \\
&&\hspace*{-1cm}  \nonumber \\
&&\hspace*{-1cm} 2{\rm T}_n(x){\rm T}_m(x) ~=~ {\rm T}_{n+m}(x) + {\rm T}%
_{n-m}(x) ~~~~~~ (n\geq m)  \,,\nonumber \\
&&\hspace*{-1cm} {\rm T}_n(1) ~=~ 1,~~~~{\rm T}_0(x) ~=~ 1, ~~~~{\rm T}%
_n(-x) ~=~ (-1)^n {\rm T}_n(x),  \label{6.2} \\
&&\hspace*{-1cm} {\rm T}_{2n}(0) ~=~ (-1)^n, ~~~~{\rm T}_{2n+1}(0) ~=~0.
\nonumber
\end{eqnarray}



\subsection{The results of calculations}

\indent

Applying formulae of the previous subsection to 
eqs.(16)-(19), we obtain the following results for eigenvalues 
(23):

\begin{eqnarray}
\chi (n,\gamma ) &=&2\Psi (1)-\Psi \Bigl(\gamma +\frac{n}{2}\Bigr)-\Psi
\Bigl%
(1-\gamma +\frac{n}{2}\Bigr)  \label{7} \\
&&  \nonumber \\
\delta (n,\gamma ) &=&-\Biggl[ \biggl(\frac{11}{3}-\frac{2}{3}\frac{n_{f}}{%
N_{c}}\biggr) \frac{1}{2}\biggl( \chi ^{2}(n,\gamma )-\Psi ^{\prime }\Bigl%
(\gamma +\frac{n}{2}\Bigr)+\Psi ^{\prime }\Bigl(1-\gamma
+\frac{n}{2}\biggr)%
\biggl)  \nonumber \\
&-&\biggl(\frac{67}{9}-2\zeta (2)-\frac{10}{9}\frac{n_{f}}{N_{c}}\biggr) %
\chi (n,\gamma ) -6\zeta (3) \nonumber \\
&+&\frac{\pi ^{2}\cos (\pi \gamma )}{\sin ^{2}(\pi \gamma
)(1-2\gamma )}\Biggl\{ \biggl(3+\biggl(1+\frac{n_{f}}{N_{c}^{3}}\biggr)
\frac{2+3\gamma (1-\gamma )}{(3-2\gamma )(1+2\gamma )}\biggr)\cdot \delta
_{n}^{0}  \nonumber \\
&-&\biggl(1+\frac{n_{f}}{N_{c}^{3}}\biggr) \frac{\gamma (1-\gamma )}{%
2(3-2\gamma )(1+2\gamma )}\cdot \delta _{n}^{2}\Biggr\}  \nonumber \\
&-&\Psi ^{\prime \prime }\Bigl(\gamma +\frac{n}{2}\Bigr)-\Psi ^{\prime
\prime }\Bigl(1-\gamma +\frac{n}{2}\biggr) +2\Phi (n,\gamma )+2\Phi
(n,1-\gamma )\Biggr],  \label{8}
\end{eqnarray}
where $\delta _{n}^{m}$ is the Kroneker symbol, and $\Psi (z)$, $\Psi
^{\prime }(z)$ and $\Psi ^{\prime \prime }(z)$ are the Euler $\Psi $
-function and its  derivatives. The function $\Phi (n,\gamma
)$ is given below
\begin{eqnarray}
\Phi (n,\gamma ) &=&-\int_{0}^{1}dx~
\frac{x^{\gamma -1 +n/2}}{1+x}\Biggr[ \frac{1%
}{2}\biggl( \Psi ^{\prime }\Bigl(\frac{n+1}{2}\Bigr)-\zeta (2)\biggr) +{\rm
Li}_{2}(-x)+{\rm Li}_{2}(x)  \nonumber \\
&+&\ln (x)\biggl(\Psi (n+1)-\Psi (1)+\ln (1+x)+\sum_{k=1}^{\infty }\frac{%
(-x)^{k}}{k+n}\biggr)  \nonumber \\
&+&\sum_{k=1}^{\infty }\frac{x^{k}}{(k+n)^{2}}(1-(-1)^{k})\Biggr]  \nonumber
\\
&&\hspace*{-1cm}\hspace{-1.3cm}=~\sum_{k=0}^{\infty }\frac{(-1)^{k}}{%
k+\gamma +n/2}\Biggl[ \Psi ^{\prime }(k+n+1)-\Psi ^{\prime }(k+1)+(-1)^{k}%
\Bigl(\beta ^{\prime }(k+n+1)+\beta ^{\prime }(k+1)\Bigr)\biggr)  \nonumber
\\
&-&\frac{1}{k+\gamma +n/2}\biggl( \Psi (k+n+1)-\Psi (k+1)%
\biggr) \Biggr],  \label{9}
\end{eqnarray}
and
\[
\beta ^{\prime }(z)=\frac{1}{4}\Biggl[ \Psi ^{\prime }\Bigl(\frac{z+1}{2}%
\Bigr)-\Psi ^{\prime }\Bigl(\frac{z}{2}\Bigr)\Biggr]
\]
For the case $n=0$ the results (\ref{8}) and (\ref{9}) coincide with ones
obtained in \cite{FL}.

Almost all terms in the right hand side of
eq.(%
\ref{8}) except the contribution
\[
\tilde \Delta
(n,\gamma )~=~\biggl(\frac{11}{3}-\frac{2}{3}\frac{n_{f}}{N_{c}%
} \biggr) \frac{1}{2}\biggl( \Psi ^{\prime }\Bigl(\gamma +\frac{n}{2}\Bigr
)-\Psi ^{\prime }\Bigl(1-\gamma +\frac{n}{2}\biggr)\biggl)
\]
are symmetric under the transformation $\gamma \leftrightarrow 1-\gamma $.
Moreover, as it was mentioned in \cite{FL}, it is possible to cancel $%
\tilde \Delta (n,\gamma )$ if one would redefine the function
$q^{\gamma -1}$
by including in it the logarithmic factor
\[
{\Biggl( \frac{\alpha _{s}(q^{2})}{\alpha _{s}(\mu ^{2})}\Biggr)}^{1/2}.
\]

Note also, that the term 
\begin{eqnarray}
\biggl(\frac{67}{9}-2\zeta (2)-\frac{10}{9}\frac{n_{f}}{N_{c}}\biggr) \chi
(n,\gamma )=\kappa \,\chi (n,\gamma)
 \label{K}
\end{eqnarray}
is proportional to the LO contribution. It appears often in the
$\overline{MS}$-scheme for the processes 
involving soft gluons and it is absent in the
gluon-bremsstrahlung (GB) scheme \cite{CaMaWe} being more natural for these
processes. In the GB-scheme the term (\ref{K}) can be incorporated to a new
coupling constant $\tilde{\alpha}_{s}(q^{2})$ as 
\begin{eqnarray}
\tilde{\alpha}_{s}(q^{2})=\alpha _{s}(q^{2}e^{-\kappa N_{c}/\beta
_{0}}))~~~~~%
\biggl(\beta _{0}=\frac{11}{3}N_{c}-\frac{2}{3}n_{f}\biggr)
 \label{K1}
\end{eqnarray}
and eq.(23) is replaced by
\begin{equation}
\omega =\frac{\tilde{\alpha}_{s}(q^{2})N_{c}}{\pi }\biggl[ \chi
(n,\gamma )+%
\tilde{\delta}(n,\gamma )\frac{\tilde{\alpha}_{s}(q^{2})N_{c}}{4\pi }\biggr]
\,,
\end{equation}
where

\begin{eqnarray}
\tilde{\delta}(n,\gamma ) &=&-\Biggl[ \biggl(\frac{11}{3}-\frac{2}{3}\frac{%
n_{f}}{N_{c}}\biggr) \frac{1}{2}\biggl( \chi ^{2}(n,\gamma )-\Psi ^{\prime
}\Bigl(\gamma +\frac{n}{2}\Bigr)+\Psi ^{\prime }\Bigl(1-\gamma +\frac{n}{2}%
\biggr)\biggl)  \nonumber \\
&-&6\zeta (3)+\frac{\pi ^{2}\cos (\pi \gamma )}{\sin ^{2}(\pi \gamma
)(1-2\gamma )}\Biggl\{ \biggl(3+\biggl(1+\frac{n_{f}}{N_{c}^{3}}\biggr)
\frac{2+3\gamma (1-\gamma )}{(3-2\gamma )(1+2\gamma )}\biggr)\cdot \delta
_{n}^{0} \nonumber \\
&-&\biggl(1+\frac{n_{f}}{N_{c}^{3}}\biggr) \frac{\gamma (1-\gamma )}{%
2(3-2\gamma )(1+2\gamma )}\cdot \delta _{n}^{2}\Biggr\}  \nonumber \\
&-&\Psi ^{\prime \prime }\Bigl(\gamma +\frac{n}{2}\Bigr)-\Psi ^{\prime
\prime }\Bigl(1-\gamma +\frac{n}{2}\biggr) +2\Phi (n,\gamma )+2\Phi
(n,1-\gamma )\Biggr].  \label{8.1}
\end{eqnarray}

Analogously  to ref.\cite{FL} we can obtain the eigenvalues (23) in the case
of a non-symmetric 
choice of 
the normalization $s_0$ of the energy in eq.(15).
For example, for the scale $s_0=q^2$, which is natural for deep-inelastic
scattering process, we have 

in $\MS$-scheme
\begin{eqnarray}
\omega =\frac{\alpha_{s}(q^{2})N_{c}}{\pi }
\Biggl[ \chi(n,\gamma )+ \biggl(
\delta(n,\gamma ) - 2 \chi(n,\gamma ) \chi'(n,\gamma )
\biggl)
\frac{\alpha_{s}(q^{2})N_{c}}{4\pi }\Biggr]
\label{8.01}
\end{eqnarray}

and in GB-scheme
\begin{eqnarray}
\omega =\frac{\tilde \alpha_{s}(q^{2})N_{c}}{\pi }
\Biggl[ \chi(n,\gamma )+ \biggl(
\tilde \delta(n,\gamma ) - 2 \chi(n,\gamma ) \chi'(n,\gamma )
\biggl)
\frac{\tilde{\alpha}_{s}(q^{2})N_{c}}{4\pi }\Biggr],
\nonumber
\end{eqnarray}
where
$$
\chi'(n,\gamma )\equiv \frac{d}{d\gamma} \chi(n,\gamma )
= -\Psi ^{\prime }\Bigl(\gamma +\frac{n}{2}\Bigr)+\Psi ^{\prime}
\Bigl(1-\gamma +\frac{n}{2}\biggr)
$$

Considering the limit $\gamma \to 0$ we obtain for $n=0$

\begin{eqnarray}
\chi(0,\gamma ) &=& \frac{1}{\gamma} + O(\gamma^2), ~~
\delta(0,\gamma ) - 2 \chi(0,\gamma ) \chi'(0,\gamma )
~=~ \frac{A}{\gamma^2} + \frac{B}{\gamma}
+ C + O(\gamma^2), \nonumber \\
\tilde \delta(0,\gamma ) &-& 2 \chi(0,\gamma ) \chi'(0,\gamma )
 ~=~ \frac{A}{\gamma^2} + 
\tilde\frac{B}{\gamma} + C + O(\gamma^2), \label{n1}
\end{eqnarray}
where

\begin{eqnarray}
A &=& -\biggl( \frac{11}{3} + \frac{2}{3}\frac{n_f}{N^3_c}
\biggr),~
B = -\frac{n_f}{9N^3_c} \left(10N_c^2 +13 \right), ~
\tilde B = -\biggl( \frac{67}{9} -2\zeta(2)+ 
\frac{13}{9}\frac{n_f}{N^3_c} \biggr), \nonumber \\
C &=&  -\biggl( \frac{395}{27} -2\zeta(3)- \frac{11}{3} \zeta(2) 
+ \frac{2n_f}{27N^3_c} \left(71 -18 \zeta(2) \right)\biggr),
\label{n2}
\end{eqnarray}

Following to ref.\cite{FL} and using eqs.(\ref{n1}) and (\ref{n2}), 
one can obtain the expression for 
anomalous dimensions of twist-2 operators $\gamma$ at $\omega \to 0$
(i.e. near $j=1$)

in $\MS$-scheme (coinciding with \cite{FL})
\begin{eqnarray}
\gamma &=&\frac{\alpha_{s}(q^{2})N_{c}}{\pi }
\biggl[ \frac{1}{\omega} +\frac{A}{4} + O(\omega)\biggr]
+
\frac{\alpha^2_{s}(q^{2})N^2_{c}}{4\pi^2 }
\biggl[ \frac{B}{\omega} + O(1) \biggr] \nonumber \\
&+&
\frac{\alpha^3_{s}(q^{2})N^3_{c}}{4\pi^3 }
\biggl[ \frac{C}{\omega^2} + O\left(\omega^{-1}\right) \biggr]
\label{n3}
\end{eqnarray}

and in GB-scheme
\begin{eqnarray}
\gamma &=&\frac{\alpha_{s}(q^{2})N_{c}}{\pi }
\biggl[ \frac{1}{\omega} +\frac{A}{4} + O(\omega)\biggr]
+
\frac{\alpha^2_{s}(q^{2})N^2_{c}}{4\pi^2 }
\biggl[ \frac{\tilde B}{\omega} + O(1) \biggr] \nonumber \\
&+&
\frac{\alpha^3_{s}(q^{2})N^3_{c}}{4\pi^3 }
\biggl[ \frac{C}{\omega^2} + O\left(\omega^{-1}\right) \biggr]
\label{n4}
\end{eqnarray}

Note, that the calculation of terms $\sim \alpha_s^3$ for anomalous
dimensions of the twist-2 operators at arbitrary values of $\omega$ will be
performed in the near future (see \cite{Verma} and references therein).

For the case $n \geq 2$ from eq.(\ref{8.01}) one can find also the 
anomalous dimensions for the local operators of the twist $t>2$.


\subsection{The asymptotics of cross-sections at $s\to \infty$}

\indent

As an example we consider
the cross-sections for the inclusive production of two pairs of charged
particles with mass $m$ in the $\gamma \gamma$ collision (see \cite{BL}):
\[
\sigma (s)~=~
\alpha^2 _{em}\alpha^2 _{s}\frac{1}{m^{2}}\frac{32}{81}\biggl(%
\sigma _{0}(s)+\Bigl(\cos ^{2}\vartheta -\frac{1}{2}
\Bigr)\sigma _{2}(s)\biggr),
\]
where $\alpha _{em}$ is electromagnetic coupling constant and 
$\sigma _{2}(s)$ and $\sigma _{0}(s)$ are the 
coefficients for angular dependent and independent contributions, 
respectively.

The asymptotic behaviour of the cross-sections $\sigma _{k}(s)~~(k=0,2)$ at
$s\to \infty $ corresponds (for the fixed coupling constant) to the unmoving 
singularities of $f_{\omega 
}(t)$ of the type $(\omega -\omega _{k})^{1/2}$, where
\[
\omega _{k}~=~ \frac{\alpha _{s} N_{c}}{\pi }\Biggl[
\chi (k,\frac{1}{2})+\frac{\alpha _{s} N_{c}}{4\pi }
\delta(k,\frac{1}{2}) \Biggr]
\equiv \frac{\alpha _{s} N_{c}}{\pi }
\chi (k,\frac{1}{2})\Biggl[1- \frac{\alpha _{s} N_{c}}{4\pi }
c(k,\frac{1}{2})\Biggr],
\]

\begin{eqnarray}
\sigma _{0}(s) &=&\frac{9\pi ^{5/2}}{32\sqrt{7\zeta (3)}}\frac{s^{\omega
_{0}}}{{\Bigl(\ln (\frac{s}{s_{0}})\Bigr)}^{1/2}}\cdot \biggl(1+O(\alpha
_{s})\biggr), \\
\sigma _{2}(s) &=&\frac{\pi ^{5/2}}{9\cdot 32\sqrt{7\zeta (3)-8}}\frac{%
s^{\omega _{2}}}{{\Bigl(\ln (\frac{s}{s_{0}})\Bigr)}^{1/2}}\cdot \biggl(%
1+O(\alpha _{s})\biggr).
\end{eqnarray}
Here the symbol $O(\alpha _{s})$ denotes unknown $\alpha _{s}$-corrections
to the formfactors.
Note, that the running of the QCD coupling constant leads to the 
substitution
of the unmoving cut by an infinite set of the Regge poles \cite{conf}.

Using our results (26), (27), we obtain the following values for $\chi (k,%
\frac{1}{2})$ and $c(k,\frac{1}{2})$ $(k=0,2)$:
\begin{eqnarray}
\chi (0,\frac{1}{2}) &=&4\ln 2,~~~
\chi (2,\frac{1}{2})~=~4(\ln2-1)\,,  \label{10} \\
&&  \nonumber \\
c(0,\frac{1}{2}) &=&2\biggl(\frac{11}{3}-\frac{2}{3}\frac{n_{f}}{N_{c}}%
\biggl) \ln 2-\frac{67}{9}+2\zeta
(2)+\frac{10}{9}\frac{n_{f}}{N_{c}}+\frac{1%
}{4\ln 2}\Biggl[ 22\zeta (3)-64{\rm Ls}_{3}\Big(\frac{\pi }{2}\Big)
\nonumber \\
&-&\frac{1}{16}\biggl(271-33\frac{n_{f}}{N_{c}^{3}}\biggr)\pi \zeta (2)%
\Biggr] ~=~25.8388+0.1869\frac{n_{f}}{N_{c}}+3.8442\frac{n_{f}}{N_{c}^{3}}\,,
\label{11} \\
&&  \nonumber \\
c(2,\frac{1}{2}) &=&2\biggl(\frac{11}{3}-\frac{2}{3}\frac{n_{f}}{N_{c}}%
\biggl)(\ln 2-1)-\frac{67}{9}+2\zeta (2)+\frac{10}{9}\frac{n_{f}}{N_{c}}
 \nonumber \\   &+&
\frac{1}{4(\ln 2-1)}  
\Biggl[ 22\zeta (3)+64{\rm Ls}_{3}\Big(\frac{\pi }{2}\Big)
+\frac{1%
}{32}\biggl(893-3\frac{n_{f}}{N_{c}^{3}}\biggr) \pi \zeta (2)-64\ln 2\Biggr]
\nonumber \\
&=&-3.2652+1.5203\frac{n_{f}}{N_{c}}+0.3947\frac{n_{f}}{N_{c}^{3}}\,,
\label{12}
\end{eqnarray}
where (see \cite{Lewin} and Appendix A)
\[
{\rm Ls}_{3}(x)=-\int_{0}^{x}\ln ^{2}\left| 2\sin \Bigl(\frac{y}{2}\Bigr%
)\right| dy \,.
\]
Note, that the function 
${\rm Ls}_{3}(x)$ appears also in  calculations of some
massive diagrams (see, for example, the recent articles \cite{LS3} and
references therein).

The LO results (33) coincide with ones obtained in \cite{BL}. The value of
$c(0,1/2)$ equals to $c(1/2)$ from \cite{FL}
(see Appendix A).

As it was discussed in \cite{FL}, the NLO correction $c(0,1/2)$ is
large and leads to a quite strong reduction of the value of the Pomeron
intercept (see recent analyses \cite{bfklp, resum} of various effective
resummations of the large NLO terms). Contrary to $c(0,1/2)$, the
correction $c(2,1/2)$ is very small and does not change the small LO value
\cite{BL} of the angle-dependent contribution.

In the GB-scheme 
we have analogously for $k=0,\,2$ (cf. (31)) 
\[
\omega _{k}~=~ \frac{\tilde \alpha _{s} N_{c}}{\pi }\Biggl[
\chi (k,\frac{1}{2})+\frac{\tilde \alpha _{s} N_{c}}{4\pi }
\tilde \delta(k,\frac{1}{2}) \Biggr]
\equiv \frac{\tilde \alpha _{s} N_{c}}{\pi }
\chi (k,\frac{1}{2})\Biggl[1- \frac{\tilde \alpha _{s} N_{c}}{4\pi }
\tilde c(k,\frac{1}{2})\Biggr]\,,
\]
where
$$
\tilde c(k,\frac{1}{2}) ~=~ c(k,\frac{1}{2}) +
\frac{67}{9}-2\zeta (2)-\frac{10}{9}\frac{n_{f}}{N_{c}}\,.
$$

Note, that in the case of the non-symmetric choice of the scale $s_0=q^2$ 
in r.h.s. of eq.(15), the values of the corrections $c(0,1/2)$
and $c(2,1/2)$ are not changed because $\chi' (k,1/2)=0$.
Thus, the asymptotics (\ref{11}) and (\ref{12}) can be applied directly 
to the deep-inelastic scattering process at small values of Bjorken variable
$x$, where $x \approx Q^2/s$.


\section{ NLO corrections to BFKL kernel in supersymmetric field theories}

\indent

\subsection{Gluino and scalar particle pair production}

As it was shown above, the eigenvalue of the BFKL kernel in the
next-to-leading approximation for QCD contains non-analytic contributions
proportional to $\delta _{n}^{0}$ and $\delta _{n}^{2}$, which in particular
leads to a violation of the holomorphic separability. It is natural to
investigate the BFKL equation in supersymmetric gauge theories where
there could be significant simplifications. In the case of the $N=1$
supersymmetric Yang-Mills theory we should consider an additional
contribution from a gluino loop. It can be obtained from the QCD result for
the next-to-leading kernel by a modification of the terms
originated from the quark loop contributions proportional to $n_{f}$. 
Such terms in $ \delta (n,\gamma )$ contain the 
factors $n_{f}/N_{c}$ or $n_{f}/N_{c}^{3}$. The
terms $\sim n_{f}/N_{c}$ are related to the quark contribution in the
running coupling constant and will be considered below. 

We discuss here only
the terms in $\delta (n,\gamma )$ proportional to $n_{f}/N_{c}^{3}$.
Their contribution to $\omega $ depends on $\gamma = 1/2+i\nu$ and
can be written as follows

\begin{equation}
\omega _{n_{f}/N_{c}^{3}}=-\frac{n_{f}}{4N_{c}}\left( \frac{\alpha _{s}}{%
\alpha }\right) ^{2}\omega _{QED}\,,
\end{equation}
where according
to eq. (B1) from Appendix B
\begin{equation}
\omega _{QED}=\frac{\alpha ^{2}\,\cos (\pi \gamma )}{\sin ^{2}(\pi \gamma
)(1-2\gamma )}\left( \frac{\left( 2+3\gamma (1-\gamma )\right) \delta
_{n}^{0}}{(3-2\gamma )(1+2\gamma )}-\frac{\gamma (1-\gamma )\,\delta
_{\left| n\right| }^{2}}{2(3-2\gamma )(1+2\gamma )}\right)
\end{equation}
coincides with the position of the Pomeranchuck singularity of the $t$%
-channel partial wave $f_{\omega }(t)$ in QED (see \cite{wu}). To give an
physical interpretation of this result one should take into account, that in
the $\omega $-plane the BFKL pomeron is the Mandelstam cut originated from
an exchange of two reggeized gluons. It means, that the amplitude for the
transition of two reggeized gluons into two others in the $t$-channel
through the fermion loop should contain the third spectral function $\rho
(s,u)
$. Therefore after the substitution $\alpha \rightarrow \alpha _{s}$ it
should coincide with the amplitude for the virtual photon-photon scattering
in QED up to the colour factor
\[
-\frac{n_{f}}{4N_{c}}=n_{f}\,\frac{1}{N_{c}^{2}-1}\,tr\,\left(
t^{a}t^{b}t^{a}t^{b}\right) \,,\,\,\left[ t^{a},t^{b}\right] =if^{abc}t^{c}
\]
for the corresponding non-planar loop diagram.

Taking into account, that gluino is a Mayorana particle contrary to the
charged electron and that its field belongs to the adjoint representation $%
(T^{a})^{bc}=-if^{abc}$ of the gauge group, we obtain for the gluino
contribution in the $N$-extended supersymmetric gauge theory the following
result
\begin{equation}
\omega _{gluino}=\frac{N}{2}\frac{N_{c}^{2}}{2}\,\left( \frac{\alpha _{s}}{%
\alpha }\right) ^{2}\omega _{QED}\,\,.
\end{equation}
Providing that $N>1$, there is also a contribution of scalar super-partners
of the gluon:

\begin{equation}
\omega _{scalar}=\frac{n_{s}(N)}{2}\frac{N_{c}^{2}}{2}\,\left( \frac{\alpha
_{s}}{\alpha }\right) ^{2}\omega _{SED}\,,
\end{equation}
where $n_{s}(N)$ is the total number of scalar particles and $\omega _{SED}$
is the position of the Pomeranchuck singularity in the scalar
electrodynamics (see eq.(B2) from Appendix B and  ref. \cite{wu}):
\begin{equation}
\omega _{SED}=\frac{\alpha ^{2}\,\cos (\pi \gamma )}{\sin ^{2}(\pi \gamma
)(1-2\gamma )}\left( \frac{\left( 1+\gamma (1-\gamma )\right) \delta
_{n}^{0}}{2(3-2\gamma )(1+2\gamma )}  +
\frac{\gamma (1-\gamma )\,\delta _{\left|
n\right| }^{2}}{4\,(3-2\gamma )\,(1+2\gamma )}\right) \,.
\end{equation}
For the number of scalar particles in the $N$-extended supersymmetric
theories we have in the case of $N=2$ and $N=4$
\begin{equation}
n_{s}(2)=2,\,\,n_{s}(4)=6\,,
\end{equation}
taking into account, that for $N=4$ the 
gluons and gluinos with both helicities
$\lambda =\pm s$ enter in one super-multiplet with the scalar 
particles. In
particular, the total contribution to $\omega $ from super-partners of the
gluon for $N=4$ is
\[
\Delta \omega _{sp}=N_{c}^{2}\,\left( \frac{\alpha _{s}}{\alpha }\right)
^{2}\left( \omega _{QED}+\frac{3}{2}\,\omega _{SED}\right) =
\]
\begin{equation}
\frac{N_{c}^{2}\,\alpha _{s}^{2}\,\cos (\pi \gamma )}{\sin ^{2}(\pi \gamma
)(1-2\gamma )}\left( \frac{\left( 11+15\,\gamma (1-\gamma )\right) \delta
_{n}^{0}}{4(3-2\gamma )(1+2\gamma )}-\frac{\gamma (1-\gamma )\,\delta
_{\left| n\right| }^{2}}{8(3-2\gamma )(1+2\gamma )}\right) \,.
\end{equation}
It is remarkable, that this term cancels the corresponding gluon
contribution non-analytic in $n$ (see (23) and (\ref{8})):
\[
\Delta \omega _{g}=-\frac{N_{c}^{2}\,\alpha _{s}^{2}\,\cos (\pi \gamma )}{%
4\sin ^{2}(\pi \gamma )(1-2\gamma )}
\left(\left(  3+\frac{\left( 2+3\gamma
(1-\gamma )\right) }{(3-2\gamma )(1+2\gamma )} \right)\delta _{n}^{0}
-\frac{\gamma
(1-\gamma )}{2(3-2\gamma )(1+2\gamma )}\delta _{\left| n\right|}^{2}\,\right).
\]

\subsection{Scalar particle contribution to the gluon Regge trajectory}

Let us consider the contribution of scalar particles to the gluon Regge 
trajectory
in the two-loop approximation. We use the results of ref. \cite{FFK}, where
the corresponding quark correction was calculated, which gives a 
possibility to find easily the gluino correction using the substitution 
$n_f \rightarrow N \,N_c$. The quark contribution can be presented in the 
form

\begin{eqnarray}
\omega^{(2)}_{n_f/N_c} (q) &=&-\frac{2n_f}{3N_c} \cdot 
\frac{\alpha^2 q^{2}}{2}
\int \frac{d^{D-2}\tilde{q}}{\tilde{q}^{2}(\tilde{q}-q)^{2}}\left[ 2\Pi
_{q}(\tilde{q})-\Pi _{q}(q)\right] ,  \label{13} \\
\Pi _{q}(q) &=&-\frac{2(1+\varepsilon )}{(3+2\varepsilon )(1+2\varepsilon
)\varepsilon }\frac{\Gamma ^{2}(1+\varepsilon )}{\Gamma (1+2\varepsilon )}%
\frac{\Gamma (1-\varepsilon )}{\left( q^{2}\right) ^{-\varepsilon }},
\nonumber
\end{eqnarray}
where $  \alpha \, \Pi _{q}(q)
\left( g^{\mu \nu }q^{2}-q^{\mu }q^{\nu }\right) $ is
one-loop quark self-energy correction to the 
polarization tensor, and $\alpha $ the is bare QCD constant
\[
\alpha =\frac{g^{2}N_{c}\Gamma (1-\varepsilon )}{(4\pi )^{2+\varepsilon }}
\]
with well-known relation between $g_{\mu }^{2}$ and $g^{2}$ in
$\overline{MS}
$ scheme: 
\begin{eqnarray}
g^{2}~=~g_{\mu }^{2}\mu ^{-2\varepsilon }\left[ 1+\frac{1}{\varepsilon }%
\left( \frac{11}{3}-\frac{2}{3}\frac{n_{f}}{N_{c}}\right) g_{\mu
}^{2}\right] ,  \label{K2}
\end{eqnarray}
where the last term in the r.h.s. of Eq.(\ref{K2}) is proportional to the
leading order term $\beta _{0}$ (\ref{K1}) of QCD $\beta $-function.

The scalar particle correction can be presented in the similar
form
\begin{eqnarray}
\omega _{sc}^{(2)}(q) &=&-\frac{2n_{s}(N)}{3N_{c}}\cdot \frac{{\overline{a}%
^{2}}q^{2}}{2}\int \frac{d^{D-2}\tilde{q}}{\tilde{q}^{2}(\tilde{q}-q)^{2}}%
\left[ 2\Pi _{sc}(\tilde{q})-\Pi _{sc}(q)\right] ,  \label{13.1} \\
\Pi _{sc}(q) &=&-\frac{1}{2(3+2\varepsilon )(1+2\varepsilon )\varepsilon }%
\frac{\Gamma ^{2}(1+\varepsilon )}{\Gamma (1+2\varepsilon )}\frac{\Gamma
(1-\varepsilon )}{\left( q^{2}\right) ^{-\varepsilon }},  \nonumber
\end{eqnarray}
where $\overline{a} \, \Pi _{sc}(q)
\left( g^{\mu \nu }q^{2}-q^{\mu }q^{\nu }\right) $ is
one-loop self-energy correction to the polarization tensor from scalar
particles and ${\overline{a}}$ is the bare constant of $N=4$ supersymmetric
theory
\[
{\overline{a}}=\frac{\overline g^{2}N_{c}}{(4\pi )^{2}}
\]
As it is well known, the coupling constant of the $N=4$ supersymmetric
theory does not run, i.e. its $\beta $-function is zero, and, hence, ${%
\overline{a}}={\overline{a}}_{\mu }$.\newline


\subsection{The results of calculations}

Putting all things together, in the case of $N=4$ supersymmetric theory we
have the following representation for the Regge trajectory
\[
\omega ^{susy}(q)=-\overline{a}_{\mu }\left( \frac{2}{\varepsilon }+2\ln
\left( \frac{q^{2}}{\mu ^{2}}\right) \right)
\]
\begin{equation}
-~\overline{a}_{\mu }^{2}\Biggl[ \left( \frac{1}{3}-2\zeta (2)\right)
\left( 
\frac{1}{\varepsilon }+2\ln \left( \frac{q^{2}}{\mu ^{2}}\right) \right) - 
\frac{8}{9}+2\zeta (3)\Biggr],
\end{equation}

Using (53), 
by analogy with the results of \cite{FL} for QCD we can easily find the
corresponding kernel $K_{r}^{susy}(q_{1},q_{2})$ for the $N=4$ SUSY 
\[
K_{r}^{susy}(q_{1},q_{2})=\frac{4\overline{a}_{\mu }\mu ^{-2\varepsilon }}{%
\pi ^{1+\varepsilon }\Gamma (1-\varepsilon )}\Biggl\{ \frac{1}{%
(q_{1}-q_{2})^{2}}\Biggl(1
\]
\[
+\overline{a}_{\mu }\Biggl[
\left( \frac{(q_{1}-q_{2})^{2}}{\mu ^{2}}\right)
{\varepsilon }\left( \frac{1}{3}-2\zeta (2)+
\varepsilon \left( -\frac{8}{9}+14\zeta (3)\right)
\right) -
{\ln }^{2}\left( {\frac{q_{1}^{2}}{q_{2}^{2}}}\right) \Biggr]
\Biggl)
\]
\[
+\overline{a}_{\mu }\Biggl[
\frac{(q_{1}^{2}-q_{2}^{2})}{(q_{1}-q_{2})^{2}(q_{1}+q_{2})^{2}}\left( \ln
\left( \frac{q_{1}^{2}}{q_{2}^{2}}\right) \ln \left( \frac{%
q_{1}^{2}q_{2}^{2}(q_{1}-q_{2})^{4}}{(q_{1}^{2}+q_{2}^{2})^{4}}\right) -2%
{\rm Li}_{2}\left( -\frac{q_{1}^{2}}{q_{2}^{2}}\right) +
2{\rm Li}_{2}\left( 
\frac{q_{1}^{2}}{q_{2}^{2}}\right) \right) 
\]
\begin{equation}
-\left( 1-\frac{(q_{1}^{2}-q_{2}^{2})}{(q_{1}-q_{2})^{2}(q_{1}+q_{2})^{2}}%
\right) \left( \int_{0}^{1}-\int_{1}^{\infty }\right) \frac{dx}{%
(q_{2}-xq_{1})^{2}}\ln \left( \frac{xq_{1}^{2}}{q_{2}^{2}}\right) \Biggr] %
\Biggr\},
\end{equation}

The set of eigenvalues for eigenfunctions (22) of the homogeneous BFKL
equation in the framework of $N=4$ supersymmetric theory 
\begin{equation}
\omega ^{susy}=
4\,\overline{a}\,
\biggl[ \chi (n,\gamma)
+\delta ^{susy}(n,\gamma )
\,\overline{a} \,
\biggr] \,
\end{equation}
can be found analogously to the previous section in the following form 
\begin{eqnarray}
 \delta ^{susy}(n,\gamma )&=&-\Biggl[ 2\Phi (n,\gamma )+2\Phi
(n,1-\gamma ) -
\biggl(\frac{1}{3}-2\zeta (2)\biggr) 
\chi (n,\gamma )  \nonumber \\
&-& 6\zeta (3)- 
 \Psi ^{\prime \prime }\Bigl(\gamma +\frac{n}{2}\Bigr)- \Psi ^{\prime
\prime } \Bigl(1-\gamma +\frac{n}{2}\biggr) \Biggr],  \label{K3}
\end{eqnarray}
where the function $\Phi (n,\gamma )$ was defined in eq.(28).

Note, that the term 
\[
\biggl(\frac{1}{3}-2\zeta (2)\biggr) \chi (n,\gamma )
\]
appears as a result of the use of the $\overline{MS}$-scheme in intermediate
calculations (see (\ref{K})). This term can be eliminated by a redefinition
of the coupling constant 
\[
{\overline{a}}\to \tilde{a}={\overline{a}}+\biggl(\frac{1}{3}-2\zeta (2)
\biggr) {\overline{a}}^2,
\]
which is equivalent to the use of the BG-scheme (see (\ref{K1})).
For the new coupling constant $\tilde{a}$ above results can be written in
the following form
\[
\omega ^{susy}=
4\, \tilde{a} \,
\biggl[ \chi (n,\gamma )+\tilde{
\delta}^{susy}(n,\gamma )
\, \tilde{a} \,
\biggr]
\,,
\]
where 
\begin{eqnarray}
\tilde{\delta}^{susy}(n,\gamma )=-\Biggl[ 2\Phi (n,\gamma )+2\Phi
(n,1-\gamma )-6\zeta (3)-\Psi ^{\prime \prime }\Bigl(\gamma
+\frac{n}{2}\Bigr%
)-\Psi ^{\prime \prime }\Bigl(1-\gamma +\frac{n}{2}\biggr) \Biggr].
 \label{K4}
\end{eqnarray}

Analogously to the previous section and ref. \cite{FL} we can obtain 
the eigenvalues of the kernel in the case
of non-symmetric 
choice of the energy-normalization $s_0$ in eq.(15).
For the scale $s_0=q^2$
we have 

in $\MS$-scheme
\begin{eqnarray}
\omega^{susy} = 4\, \overline{a}\,
\Biggl[ \chi(n,\gamma )+ \biggl(
\delta^{susy}(n,\gamma ) - 2 \chi(n,\gamma ) \chi'(n,\gamma )
\biggl)
\, \overline{a} \,
\Biggr]
\label{K4.1}
\end{eqnarray}

and in GB-scheme
\begin{eqnarray}
\omega^{susy} = 4\, \tilde{a} \,
\Biggl[ \chi(n,\gamma )+ \biggl(
\tilde \delta^{susy}(n,\gamma ) - 2 \chi(n,\gamma ) \chi'(n,\gamma )
\biggl)
\, \tilde{a} \,
\Biggr].
\nonumber
\end{eqnarray}
The 
anomalous dimensions $\gamma^{susy}$ of the twist-2 operators
at $\omega \to 0$ (and $n=0$) are

in $\MS$-scheme
\begin{eqnarray}
\gamma^{susy} = 4\, \overline{a}\, \Biggl[
\biggl( \frac{1}{\omega} + O(\omega)\biggr)
+ \overline a \,
\biggl( \frac{B^{susy}}{\omega} + O(1) \biggr)
+ \overline{a}^2 \,
\biggl( \frac{C^{susy}}{\omega^2} + O\left(\omega^{-1}\right) \biggr)
\Biggl]
\label{n5}
\end{eqnarray}

and in GB-scheme
\begin{eqnarray}
\gamma^{susy} =  4\, \tilde{a} \, \Biggl[
\biggl( \frac{1}{\omega} + O(\omega)\biggr)
+ \tilde{a} \,
\biggl( \frac{\tilde B^{susy}}{\omega} + O(1) \biggr)
+ \tilde{a}^2 \,
\biggl( \frac{C^{susy}}{\omega^2} + O\left(\omega^{-1}\right) \biggr)
\Biggl],
\label{n6}
\end{eqnarray}
where

\begin{eqnarray}
B^{susy} ~=~ \frac{1}{3} ,~~~
\tilde B^{susy} ~=~ 2\zeta(2), ~~~\mbox{ and }~~~
C^{susy} ~=~  2\zeta(3).
\label{n7}
\end{eqnarray}

For the case $n \geq 2$ one can calculate also the anomalous dimensions 
of the corresponding local operators. Because in the case $N=4$ SUSY the 
result is analytic in $|n|$, one can continue these anomalous dimensions 
to the negative values of $|n|$. It gives a possibility to find the singular
contributions of the anomalous dimensions of the twist-2 operators not 
only at $j=1$ but also at 
other integer points $j=0,\,-1,\,-2...$. In particular, in the Born 
approximation for the anomalous dimension of the operator corresponding 
to the sum of distributions of gluons, gluino and scalar particles we 
obtain $\gamma = 4\, \overline{a}\,
(\Psi (1)-\Psi (j-1))$ 
which coincides with the result of the direct calculations (see 
\cite{N=0}). Thus, in the case $N=4$ the BFKL equation presumably contains 
the information sufficient for restoring the kernel of the
DGLAP equation.


\section{ Conclusion}

In this paper we calculated next-to-leading corrections to eigenvalues of
the BFKL kernel in QCD for arbitrary conformal spins $n$ (see (\ref{8})).
Now with the use of eqs. (39), (42) one can find 
the spin-dependent contribution to
the total cross-section for the $\gamma ^* \gamma ^*$  annihilation into 
hadrons at 
high energies. We calculated also next-to-leading corrections to the BFKL
equation in the supersymmetric field theories and
 next-to-leading corrections to anomalous dimensions of 
twist-2 operators (see (56), (59)).
In a particular case of the
extended $N=4$ SUSY the result of our calculations is significantly
simplified (see eqs. (\ref{K3}) and (\ref{K4})). 
The absence of the coupling
constant renormalization in this model leads presumable to the M\"{o}bius
invariance of the BFKL equation in higher orders of the perturbation theory.
This invariance can allow one to guess the next-to-leading corrections to the
BFKL kernel in QCD\ for a general case $t\neq 0$, which is very important
for the theory of the processes at small $x$. The remarkable cancellation of
non-analytic contributions proportional to $\delta _{n}^{0}$ and $\delta
_{n}^{2}$ at $N=4$ SUSY is a possible manifestation of the integrability of
the reggeon dynamics in the Maldacena model \cite{Malda} corresponding to
the limit $N_{c}\rightarrow \infty $. Note, that in this model the 
eigenvalues of the LLA pair kernels in the evolution equations for the matrix
elements of the quasi-partonic operators are proportional to $\psi
(j-1)-\psi (1)$ \cite{N=0}, which means, that the corresponding Hamiltonian
coincides with the local Hamiltonian for the integrable Heisenberg spin
model.
The residues of these eigenvalues in the points $j=-k$ can be obtained 
from the BFKL equation by the analytic continuation of the anomalous 
dimensions to negative integer values of the conformal spin $|n|$. 
Therefore the DGLAP equation is not independent from the BFKL equation in 
$N=4$
SUSY and their integrability properties at $N_c \rightarrow \infty $ are 
related. We shall return to these problems in our future publications.

\vspace{1cm} \hspace{1cm} {\Large {} {\bf Acknowledgments} 
}

A. Kotikov thanks the Alexander von Humboldt Foundation for its support. L.
Lipatov thanks the Hamburg University for
its hospitality during the period of time when this work was done. 
He is supported  by the
INTAS grant 97-31696 and by the Deutsche Forschungsgemeinschaft.

We are indebted to J. Bartels, V. Fadin, V. Kim, R. Kirschner, R.
Peschansky and participants of the PNPI Winter School and of the Gribov-70
Workshop for helpful discussions.




\setcounter{secnumdepth}{2}
\addcontentsline{toc}{section}{APPENDIX}

\setcounter{section}{0}
\setcounter{subsection}{0}
\setcounter{equation}{0}
 \def\thesection{\Alph{section}}
\def\thesubsection {\thesection.\arabic{subsection}}
\def\theequation{\thesection.\arabic{equation}}

\section{Appendix}

Here we demonstrate how the function ${\rm Ls}_{3}(x)$ appears in
eqs.(\ref{11}) and (\ref{12}).

The direct evaluation of the values of $c(0,1/2)$ and $c(2,1/2)$ leads to
\begin{eqnarray}
c(0,\frac{1}{2}) &=&2\biggl(\frac{11}{3}-\frac{2}{3}\frac{n_{f}}{N_{c}}%
\biggl) \ln 2-\frac{67}{9}+2\zeta
(2)+\frac{10}{9}\frac{n_{f}}{N_{c}}+\frac{1%
}{4\ln 2}\Biggl[ 22\zeta (3)  \nonumber \\
&+&16\int_{0}^{1}\frac{dt}{t}\arctan (\sqrt{t})\ln \left( \frac{1}{1-t}%
\right) +\frac{1}{16}\biggl(81+33\frac{n_{f}}{N_{c}^{3}}\biggr)\pi \zeta
(2)%
\Biggr]  \label{A1} \\
&&  \nonumber \\
c(2,\frac{1}{2}) &=&2\biggl(\frac{11}{3}-\frac{2}{3}\frac{n_{f}}{N_{c}}%
\biggl)(\ln 2-1)-\frac{67}{9}+2\zeta (2)+\frac{10}{9}\frac{n_{f}}{N_{c}}+%
\frac{1}{4(\ln 2-1)}\Biggl[ 22\zeta (3)  \nonumber \\
&-&16\int_{0}^{1}\frac{dt}{t}\arctan (\sqrt{t})\ln \left( \frac{1}{1-t}%
\right) +\frac{1}{32}\biggl(189-3\frac{n_{f}}{N_{c}^{3}}\biggr) \pi \zeta
(2)-64\ln 2\Biggr]  \label{A2}
\end{eqnarray}
Note that (\ref{A1}) coincides exactly with eq.(16) of \cite{FL}.

To obtain eqs. (\ref{11}) and (\ref{12}) we should evaluate the integral
\[
\int_{0}^{1}\frac{dt}{t}\arctan (\sqrt{t})\ln \left( \frac{1}{1-t}\right)
\]
Using the following substitutions
\begin{eqnarray}
t=p^{2},~~~
\arctan (x+iy)=\frac{i}{2}\ln \frac{1+y-ix}{1-y+ix}+2\pi
k~~~(i=e^{i\pi /2})\,,  \nonumber
\end{eqnarray}
we obtain for it the following representation (for $k=0$)

\begin{eqnarray}
\int^1_0 \frac{dt}{t} \arctan (\sqrt{t})\ln \left(\frac{1}{1-t}\right) &=&
-i \int^1_0 \frac{dp}{p} \biggl( \ln (1-p) + \ln (1+p) \biggl)  \nonumber \\
\cdot \biggl( \ln (1-ip) - \ln (1+ip) \biggl) &=& i \biggl[ {\rm F}_1(i) +
{\rm F}_2(i) -{\rm F}_1(-i) - {\rm F}_2(-i) \biggl],  \label{A3}
\end{eqnarray}
where \cite{Devoto}

\begin{eqnarray}
{\rm F}_1(a) &\equiv& \int^1_0 \frac{dp}{p} \ln (1-p)\ln (1+ap) = {\rm Li}%
_3(-a) + {\rm S}_{1,2}(-a) ~~~~~~~\mbox{ and}  \nonumber \\
{\rm F}_2(a) &\equiv& \int^1_0 \frac{dp}{p} \ln (1+p) \ln (1+ap) = {\rm S}%
_{1,2}\left(\frac{1-a}{1+a}\right) - {\rm S}_{1,2}\left(\frac{a-1}{1+a}%
\right) + {\rm S}_{1,2}(-a)  \nonumber \\
&+& {\rm Li}_3(a) - \frac{7}{8} \zeta(3) +\frac{1}{2}\ln^2a \ln
\left(\frac{%
1-a}{1+a}\right) - \ln a \biggl[ \zeta(2) - {\rm Li}_2
\left(\frac{1-a}{1+a}%
\right) \biggr]  \label{A4}
\end{eqnarray}
Here ${\rm S}_{n,p}(x)$ are the Nielsen generalized polylogarithms (see
\cite
{KoMiRe, Devoto}):

\begin{eqnarray}
{\rm S}_{n,p}(x) ~=~ \frac{(-1)^{n+p-1}}{(n-1)!p!} \int^1_0 \frac{dy}{y}
\ln^{n-1} y \ln (1-xy)  \label{A5}
\end{eqnarray}
and the polylogarithms ${\rm Li}_n(x) = {\rm S}_{n-1,1}(x)$.

With the use of (\ref{A4}) we have
\begin{eqnarray}
{\rm F}_{1}(\pm i) &=&{\rm Li}_{3}(\mp i)+{\rm S}_{1,2}(\mp i),
\nonumber \\
{\rm F}_{2}(\pm i) &=&2{\rm S}_{1,2}(\mp i)-{\rm S}_{1,2}(\pm i)+{\rm Li}%
_{3}(\pm i)-\frac{7}{8}\zeta (3)\pm \frac{\pi }{2}i\biggl[ {\rm Li}_{2}(\mp
i)+\frac{1}{2}\zeta (2)\biggr] \,. \label{A6}
\end{eqnarray}

Using the integral representation (\ref{A5}), we have for the Euler
dilogarithm ${\rm Li}_2(i)$:
\[
{\rm Li}_2(1) - {\rm Li}_2(i) ~=~ \int_1^i \frac{dp}{p} \ln (1-p) \,.
\]
If we replace $s$ by $s= e^{i\theta}$    corresponding to 
\[
1-s= 1-e^{i\theta}=2\sin
\left(\frac{\theta}{2} \right) e^{i(\theta -\pi)/2}\,,
\]
we obtain
\[
{\rm Li}_2(1) - {\rm Li}_2(i) ~=~ 
\frac{9}{8} \zeta(2) - i{\rm G},
\]
where
\[
{\rm G} \equiv -\int^1_0 dx \frac{\ln x}{1+x^{2}} ={\rm Cl}_2(\pi/2).
\]
Here  according to ref. \cite{Lewin}
\[
{\rm Cl}_2(x)= {\rm Ls}_2(x)
~~\mbox{ and }~~ {\rm Ls}_n(x)=-\int_0^x \ln^{n-1}\left|2 \sin
\Bigl(\frac{y%
}{2} \Bigr)\right| dy.
\]

Thus,
\begin{eqnarray}
{\rm Li}_{2}(\pm i) ~=~ -\frac{1}{8} \zeta(2) \pm i{\rm G}.  \label{A7}
\end{eqnarray}

Repeating above calculations, one can derive the relations
\begin{eqnarray}
{\rm Li}_{3}(\pm i) &=& -\frac{3}{32} \zeta(3) \pm \frac{3}{16} i \pi
\zeta(2),  \label{A8} \\
{\rm S}_{1,2}(\pm i) &=& \frac{29}{64} \zeta(3) - \frac{\pi}{4} {\rm G} \mp
\frac{i}{2} \biggl[ {\rm Ls}_3\left(\frac{\pi}{2} \right) + \frac{7}{16} \pi
\zeta(2) \biggr].  \label{A9}
\end{eqnarray}
The obtained results (\ref{A7})-(\ref{A9}) are in agreement with ref.
\cite{Lewin}.

Putting (\ref{A7})-(\ref{A9}) to (\ref{A6}) and (\ref{A3}), we have
\[
\int_{0}^{1}\frac{dt}{t}\arctan (\sqrt{t})\ln \left( \frac{1}{1-t}\right)
~=~-4{\rm Ls}_{3}\left( \frac{\pi }{2}\right) -\frac{11}{8}\pi \zeta (2)
\]
and, thus, eqs.(\ref{A1}) and (\ref{A2}) coincide with Eqs.(\ref{11}) and (%
\ref{12}), respectively.

\setcounter{equation}{0}

\section{Appendix}

Although the Pomeranchuck singularity in QED was investigated many years ago
(see \cite{wu}), here we review the obtained results and generalize them to
the case of the charged scalar particle production.

To begin with, let us consider the electron-electron scattering in the Regge
kinematics $2p_{A}p_{B}=s\gg q^{2}$, where $p_{A}$ and $p_{B}$ are momenta
of the colliding electrons. In the Born approximation the corresponding
amplitude equals

\[
A_{ee\rightarrow ee}=j_{\mu }(p_{A})\,\frac{\delta ^{\mu \nu }}{q^{2}}j_{\nu
}(p_{B})=j_{\mu }(p_{A})\,p_{B}^{\mu }\frac{2/s}{q^{2}}j_{\nu
}(p_{B})p_{A}^{\nu }=2e^{2}\frac{s}{q^{2}}\,,\,
\]
where $j_{\mu }(p)=e\,\overline{u}(p^{\prime })\,\gamma _{\mu }\,u(p)$. 
The elastic cross-section is

\[
\sigma _{el}=\frac{1}{4}\int \frac{d^{2}q}{(2\pi )^{2}}\frac{\left|
A_{ee\rightarrow ee}\right| ^{2}}{s^{2}}=4\alpha ^{2}\int \frac{d^{2}q}{%
\left| q\right| ^{4}}\,.
\]
This cross-section is divergent at small $q$ due to an infinite radius of
the Coulomb interaction. We shall ignore this circumstance because in the
case of the scattering of non-charged particles $A$ and $\,B$ the integrand
is multiplied by a product of their impact factors $\Phi _{A}(q)$ and $%
\Phi _{B}(q)$ vanishing at $q=0$

\[
\sigma _{AB}=\int \frac{d^{2}q}{(2\pi )^{2}\left| q\right| ^{4}}\,\Phi
_{A}(q)\,\Phi _{B}(q)\,\,,
\]
which leads to the cancellation of this divergency (see \cite{wu}). Note, 
that for 
the electron being a charged particle the impact factor is non-zero for 
$q=0$: \[
\Phi _{e}(q)=4\pi \alpha
\]

The amplitude for the production of a massless electron-positron pair in the
$ee$ scattering is

\[
A_{ee\rightarrow eee^{+}e}=2e^{4}\frac{s}{q_{1}^{2}q_{2}^{2}}A_{\gamma
*\gamma *\rightarrow e^{+}e}\,,
\]
where the factor $A_{\gamma *\gamma *\rightarrow e^{+}e}$ describes the
transition of two virtual photons into the $e^{+}e^{-}$ pair 

\[
A_{\gamma *\gamma *\rightarrow e^{+}e^{-}}=\overline{u}_{e^{-}}(k_{1})%
\,A_{e}\,v_{e^{+}}(-k_{2})\,,\,\,q_{1}-q_{2}=k_{1}+k_{2}\,.
\]
The amplitude $A_{e}$ is given below

\[
A_{e}=\frac{2}{s}\left( \widehat{p_{A}}\frac{\widehat{k_{1}^{\perp }}-%
\widehat{q_{1}^{\perp }}}{\left| k_{1}-q_{1}\right| ^{2}+\left| k_{1}\right|
^{2}\frac{\beta _{2}}{\beta _{1}}}\widehat{p_{B}}+\widehat{p_{B}}\frac{-%
\widehat{k_{2}^{\perp }}+\widehat{q_{1}^{\perp }}}{\left| k_{2}-q_{1}\right|
^{2}+\left| k_{2}\right| ^{2}\frac{\beta _{1}}{\beta _{2}}}\widehat{p_{A}}%
\right) ,
\]
where we introduced the Sudakov variables for momenta $k_{1}$ and $k_{2}$ of
the produced electron and positron

\[
k_{r}=\alpha \,_{r}p_{B}+\beta _{r}p_{A}+k_{r}^{\perp }\,,\,\,s\alpha
_{r}\beta _{r}=-(k_{r}^{\perp })^{2}=\left| k_{r}\right| ^{2}
\]
and used for the momenta of the virtual photons the following relations

\[
q_{1}=(\beta _{1}+\beta _{2})\,p_{A}+q_{1}^{\perp }\,,\,\,q_{2}=-(\alpha
_{1}+\alpha _{2})\,p_{B}+q_{2}^{\perp }
\]
valid at high energies when $2k_{1}k_{2}\ll s$ (corresponding to small $%
\alpha _{r}$ and $\beta _{r}$). Note, that with the use of the Dirac
equation for the produced particles we can rewrite $A_{e}$ in the other form
\[
A_{e}\rightarrow \frac{2}{s(\beta _{1}+\beta _{2})}\left( \frac{s\,\beta
_{2}\,\widehat{q_{1}^{\perp }}+\,\widehat{q_{1}^{\perp }}\,\,\widehat{%
q_{2}^{\perp }}\,\,\widehat{p_{B}}}{\left| k_{1}-q_{1}\right| ^{2}+\left|
k_{1}\right| ^{2}\frac{\beta _{2}}{\beta _{1}}}-\frac{s\,\beta _{1}\,%
\widehat{q_{1}^{\perp }}+\,\widehat{p_{B}}\,\widehat{q_{2}^{\perp }}\,\,%
\widehat{q_{1}^{\perp }}\,\,}{\left| k_{1}+q_{2}\right| ^{2}+\left|
k_{1}-q_{1}+q_{2}\right| ^{2}\frac{\beta _{1}}{\beta _{2}}}\right)
\]
and verify its vanishing at $q_{1}^{\perp }=0$ or $q_{2}^{\perp }=0$.

Analogously for the production of a pair of charged scalar particles $s$ we
have

\[
A_{ee\rightarrow ees^{+}s^{-}}=2e^{4}\frac{s}{q_{1}^{2}q_{2}^{2}}A_{s}\,,
\]
where
\[
A_{s}=-2+\frac{2\,\left| k_{1}\right| ^{2}}{\left| k_{1}-q_{1}\right|
^{2}+\left| k_{1}\right| ^{2}\frac{\beta _{2}}{\beta _{1}}}+\frac{2\,\left|
k_{2}\right| ^{2}}{\left| k_{2}-q_{1}\right| ^{2}+\left| k_{2}\right| ^{2}%
\frac{\beta _{1}}{\beta _{2}}}\,.
\]
It can be written in the form

\[
A_{s}\rightarrow \frac{2}{\beta _{1}+\beta _{2}}\,\,\left( \frac{\beta
_{2}(%
\overrightarrow{q_{1}},2\overrightarrow{k_{1}}-\overrightarrow{q_{1}})}{%
\left| k_{1}-q_{1}\right| ^{2}+\left| k_{1}\right| ^{2}\frac{\beta _{2}}{%
\beta _{1}}}+\frac{\beta
_{1}\,(\overrightarrow{q_{1}},2\overrightarrow{k_{2}%
}-\overrightarrow{q_{1}})}{\left| k_{2}-q_{1}\right| ^{2}+\left|
k_{2}\right| ^{2}\frac{\beta _{1}}{\beta _{2}}}\right) \,
\]
to make obvious its vanishing at $q_{1}=0$ or at $q_{2}=0$. The total
cross-section for the pair production in the collision of the particles $A$
and $B$ after calculating the integrals over $\alpha _{i}$ ($i=1,2$) with
the use of the $\delta $-functions $\delta (s\alpha _{i}\beta _{i}-\left|
k_{i}\right| ^{2})$ can be presented as follows

\[
\sigma _{pair}=\log \frac{s}{m^{2}}\,\int \frac{d^{2}q_{1}\,\,\Phi
_{A}(q_{1})}{(2\pi )^{2}\,\left| q_{1}\right| ^{4}\,}\,\frac{%
d^{2}q_{1}\,\Phi _{B}(q_{2})}{(2\pi )^{2}\,\left| q_{2}\right| ^{4}}\,K(%
\overrightarrow{q_{1}},\overrightarrow{q_{2}})\,.
\]
The kernel $K(\overrightarrow{q_{1}},\overrightarrow{q_{2}})$ for
electrons $e$ and scalar particles $s$ is correspondingly

\[
K_{e,s}(\overrightarrow{q_{1}},\overrightarrow{q_{2}})=16\alpha ^{2}\int
\frac{d^{2}k_{1}}{(2\pi )^{2}\,}\int_{0}^{1}dx\,\,\,f_{e,s}(\overrightarrow{
k_{1}};x)\,,\,\,x=\frac{\beta _{1}}{\beta _{1}+\beta _{2}}
\]
and
\[
f_{e}(\overrightarrow{k_{1}};x)=\frac{tr(\widehat{k_{1}}A_{e}\widehat{k_{2}}
A_{e}^{+})}{16\,x(1-x)}\,\,,\,\,\,\,f_{s}(\overrightarrow{k_{1}};x)=\frac{%
\left| A_{s}\right| ^{2}}{16\,x(1-x)}\,\,.
\]
The factor $\log (s/m^2)
$ was obtained as a result of the
integration over $\beta =\beta _{1}+\beta _{2}$ in the region $m^{2}/s\ll
\beta \ll 1$, where the mass scale $m$ equals to an essential value for
transverse momenta of charged particles inside the colliding particles. We
obtain for the $e^{+}e^{-}$ production
\[
f\,_{e}(\overrightarrow{k_{1}};x)=\frac{\frac{1}{2}\left| k_{1}\right|
^{2}\left| k_{1}-q_{1}\right| ^{2}}{\left( \left| k_{1}-xq_{1}\right|
^{2}+x(1-x)\left| q_{1}\right| ^{2}\right) ^{2}}+\frac{\frac{1}{2}\left|
k_{2}\right| ^{2}\left| k_{1}+q_{2}\right| ^{2}}{\left( \left|
k_{1}+q_{2}-xq_{1}\right| ^{2}+x(1-x)\left| q_{2}\right| ^{2}\right) ^{2}}-
\]
\[
\frac{(\overrightarrow{k_{1}}-\overrightarrow{q_{1}},\overrightarrow{q_{2}}+
\overrightarrow{k_{1}})\,(\overrightarrow{k_{2}},\overrightarrow{k_{1}})-(%
\overrightarrow{k_{1}}-\overrightarrow{q_{1}},\overrightarrow{k_{2}})\,(%
\overrightarrow{k_{1}}+\overrightarrow{q_{2}},\overrightarrow{k_{1}})-(%
\overrightarrow{k_{1}}+\overrightarrow{q_{2}},\overrightarrow{k_{2}})\,(%
\overrightarrow{k_{1}}-\overrightarrow{q_{1}},\overrightarrow{k_{1}})}{%
\left( \left| k_{1}-xq_{1}\right| ^{2}+x(1-x)\left| q_{1}\right| ^{2}\right)
\left( \left| k_{1}+q_{2}-xq_{1}\right| ^{2}+x(1-x)\left| q_{1}\right|
^{2}\right) }
\]
and for charged scalar pair production
\[
f\,_{s}(\overrightarrow{k_{1}};x)=\frac{\frac{1}{4}\,x(1-x)\,(%
\overrightarrow{q_{1}},2\overrightarrow{k_{1}}-\overrightarrow{q_{1}})^{2}}{
\left( \left| k_{1}-xq_{1}\right| ^{2}+x(1-x)\left| q_{1}\right| ^{2}\right)
^{2}}+\frac{\frac{1}{4}\,x(1-x)\,(\overrightarrow{q_{1}},2\overrightarrow{%
k_{2}}-\overrightarrow{q_{1}})^{2}}{\left( \left| k_{1}+q_{2}-xq_{1}\right|
^{2}+x(1-x)\left| q_{2}\right| ^{2}\right) ^{2}}+
\]
\[
\frac{\frac{1}{2}\,x(1-x)\,(\overrightarrow{q_{1}},2\overrightarrow{k_{1}}-%
\overrightarrow{q_{1}})\,\,(\overrightarrow{q_{1}},2\overrightarrow{k_{2}}-%
\overrightarrow{q_{1}})}{\left( \left| k_{1}-xq_{1}\right| ^{2}+x(1-x)\left|
q_{1}\right| ^{2}\right) \left( \left| k_{1}+q_{2}-xq_{1}\right|
^{2}+x(1-x)\left| q_{1}\right| ^{2}\right) }\,.
\]

By introducing the Feynman parameter $y$ to unify two factors in the
denominators of the last contributions to
$f\,_{e}(\overrightarrow{k_{1}};x)$
and $f\,_{s}(\overrightarrow{k_{1}};x)$ and integrating the kernel over $%
\overrightarrow{k_{1}}$ one can present it in the form (cf. \cite{wu})

\[
K_{e,s}(\overrightarrow{q_{1}},\overrightarrow{q_{2}})=\frac{8\alpha ^{2}}{%
\pi }\int_{0}^{1}dx\int_{0}^{1}dy\,\,\,\varphi
_{e,s}(\overrightarrow{q_{1}},%
\overrightarrow{q_{2}};x,y)\,,
\]
where for the electrons
\[
\varphi _{e}(\overrightarrow{q_{1}},\overrightarrow{q_{2}};x,y)=
\]
\[
\frac{\left( x(1-x)+x(1-y)-4x(1-x)y(1-y)\right) \left| q_{1}\right|
^{2}\left| q_{2}\right| ^{2}-2x(1-x)y(1-y)\left( \overrightarrow{q_{1}},%
\overrightarrow{q_{2}}\right) ^{2}}{x(1-x)\left| q_{1}\right|
^{2}+y(1-y)\left| q_{2}\right| ^{2}}
\]
and for scalar particles

\[
\varphi _{s}(\overrightarrow{q_{1}},\overrightarrow{q_{2}};x,y)=
\]
\[
\frac{\left( x(1-x)+x(1-y)-8x(1-x)y(1-y)\right) \left| q_{1}\right|
^{2}\left| q_{2}\right| ^{2}+4x(1-x)y(1-y)\left( \overrightarrow{q_{1}},%
\overrightarrow{q_{2}}\right) ^{2}}{4\left( x(1-x)\left| q_{1}\right|
^{2}+y(1-y)\left| q_{2}\right| ^{2}\right) }.
\]
Note, that for the supersymmetric QED, where there are scalar charged 
particles, the total contribution to the kernel is \[
\frac{\varphi _{e}(\overrightarrow{q_{1}},\overrightarrow{q_{2}}%
;x,y)+2\varphi _{s}(\overrightarrow{q_{1}},\overrightarrow{q_{2}};x,y)}{%
\left| q_{1}\right| ^{2}\left| q_{2}\right| ^{2}}=\frac{%
3x(1-x)+3x(1-y)-16x(1-x)y(1-y)}{2\left( x(1-x)\left| q_{1}\right|
^{2}+y(1-y)\left| q_{2}\right| ^{2}\right) }
\]
and does not contain the term proportional to
$\left( \overrightarrow{q_{1}},%
\overrightarrow{q_{2}}\right) ^{2}$.

The total cross-section in LLA can be written in the form
\[
\sigma _{tot}=\int_{a-i\infty }^{a+i\infty }\frac{d\omega }{2\pi }
\left( 
\frac{s}{m^{2}}\right) ^{\omega }\,\int \frac{d^{2}q_{1}\,\,\Phi _{A}(q_{1})%
}{(2\pi )^{2}\,\left| q_{1}\right| ^{2}\,}\,\frac{d^{2}q_{1}\,\Phi
_{B}(q_{2})}{(2\pi )^{2}\,\left| q_{2}\right| ^{2}}\,f_{\omega }(%
\overrightarrow{q_{1}},\overrightarrow{q_{2}})\,,
\]
where the $t$-channel partial wave $f_{\omega }(\overrightarrow{q_{1}},%
\overrightarrow{q_{2}})$ for the $\gamma ^{*}\gamma ^{*}$-scattering
satisfies the Bethe-Salpeter equation

\[
\omega f_{\omega }(\overrightarrow{q_{1}},\overrightarrow{q_{2}})=(2\pi
)^{2}\delta ^{2}(q-q^{\prime })+\int \frac{d^{2}q_{1}^{\prime }\,\,}{(2\pi
)^{2}\,\left| q_{1}\right| ^{2}\,\left| q_{1}^{\prime }\right| ^{2}}\,K(%
\overrightarrow{q_{1}},\overrightarrow{q_{1}^{\prime }})\,\,f_{\omega }(%
\overrightarrow{q_{1}^{\prime }},\overrightarrow{q_{2}})\,.
\]

The solution of this equation can be obtained in the form of the Fourier
expansion
\[
f_{\omega }(\overrightarrow{q_{1}},\overrightarrow{q_{2}})\,=\sum_{n=-\infty
}^{\infty }\int_{-\infty }^{\infty }d\,\nu \,\,e^{i\,n\,(\varphi
_{1}-\varphi _{2})}\,\frac{2\pi }{\left| q_{1}\right| \left| q_{2}\right| }%
\left( \frac{\left| q_{1}\right| }{\left| q_{2}\right| }\right) ^{2i\,\nu }%
\frac{1}{\omega -\omega (n,\,\nu )}\,,
\]
where $\varphi _{k}$ is the phase of the two-dimensional vector $%
\overrightarrow{q_{k}}$ written in the complex form as $\left| q_{k}\right|
e^{i\,\varphi _{k}}$ and $\omega (n,\,\nu )$ is the eigenvalue of the
homogeneous Bethe-Salpeter equation:
\[
\omega (n,\,\nu )=\int \frac{d^{2}q^{\prime }}{(2\pi )^{2}\left| q\right|
^{4}}\,K(\overrightarrow{q},\,\overrightarrow{q^{\prime }}%
)\,e^{i\,n\,(\varphi ^{\prime }-\varphi )}\left( \frac{\left| q^{\prime
}\right| }{\left| q\right| }\right) ^{2i\nu -3}.
\]

Using the above expressions for $K_{e,\,s}(\overrightarrow{q},\,%
\overrightarrow{q^{\prime }})$, it is easy to calculate $\omega (n,\nu )$
for the spinor and scalar electrodynamics (cf. \cite{wu})
\begin{eqnarray}
\omega _{QED}(n,\,\nu ) &=&\frac{\alpha ^{2}}{32}\,\frac{\sinh (\pi \nu )}{%
\left( \cosh (\pi \nu )\right) ^{2}}\,\frac{1}{1+\nu ^{2}}\,\frac{1}{\nu }%
\,\left( (11+12\,\nu ^{2})\,\delta _{n}^{0}-\frac{1+4\nu ^{2}}{2}\,\delta
_{\left| n\right| }^{2}\right) , \\
\omega _{SED}(n,\,\nu ) &=&\frac{\alpha ^{2}}{64}\,\frac{\sinh (\pi \nu )}{%
\left( \cosh (\pi \nu )\right) ^{2}}\,\frac{1}{1+\nu ^{2}}\,\frac{1}{\nu }%
\,\left( (5+4\,\nu ^{2})\,\delta _{n}^{0}+\frac{1+4\nu ^{2}}{2}\,\delta
_{\left| n\right| }^{2}\right) .
\end{eqnarray}
The high energy behaviour of the total cross-section is
\[
\sigma _{tot}\sim \sum_{n=0}^{\infty }c_{n}\,s^{\Delta (n)}\,\cos (n(\varphi
^{\prime }-\varphi ))
\]
where $\Delta $ is the Pomeron intercept:
\begin{eqnarray}
\Delta _{QED}(n) &=&\pi \alpha ^{2}\,\left( \frac{11}{32}\,\delta _{n}^{0}-%
\frac{1}{64}\,\delta _{n}^{2}\right)  \\
\Delta _{SED}(n) &=&\pi \alpha ^{2}\,\left( \frac{5}{64}\,\delta _{n}^{0}+%
\frac{1}{128}\,\delta _{n}^{2}\right) .
\end{eqnarray}
Note, that for the supersymmetric QED the results are especially simple:

\[
\omega _{SUSY}(n,\,\nu )=\frac{\alpha ^{2}}{2\nu }\,\frac{\sinh (\pi
\nu )}{%
\left( \cosh (\pi \nu )\right) ^{2}}\,\,\,\delta _{n}^{0}\,,\,\,\Delta
_{SUSY}(n)=\frac{\pi \alpha ^{2}}{2}\delta _{n}^{0}\,.
\]

\indent



\end{document}